\documentclass[sigconf]{acmart}

\usepackage{booktabs}

\usepackage[normalem]{ulem}  
\usepackage{bbold}  
\usepackage{subfigure}  
\usepackage{caption} 
\usepackage{placeins}  
\usepackage{multirow} 
\usepackage{moreverb} 
\usepackage{enumerate} 
\usepackage{enumitem}  
\usepackage{tabularx} 
\usepackage{url}

\usepackage{tikz}
\usepackage{pgfplots}
\usetikzlibrary{arrows}
\usepackage{filecontents}
\usepackage{pgfplots,pgfplotstable}

\newcommand{\vect}[1]{\boldsymbol{#1}}

\newcommand{\subsectionshrink}{\vspace*{-0.5\baselineskip}}

\begin{document}

\copyrightyear{2018} 
\acmYear{2018} 
\setcopyright{acmcopyright}
\acmConference[CIKM '18]{The 27th ACM International Conference on Information and Knowledge Management}{October 22--26, 2018}{Torino, Italy}
\acmBooktitle{The 27th ACM International Conference on Information and Knowledge Management (CIKM '18), October 22--26, 2018, Torino, Italy}
\acmPrice{15.00}
\acmDOI{10.1145/3269206.3269257}
\acmISBN{978-1-4503-6014-2/18/10}

\title{IntentsKB: A Knowledge Base of Entity-Oriented Search Intents}

%
\author{Dar\'{i}o Garigliotti}
\affiliation{University of Stavanger}
\email{dario.garigliotti@uis.no}

\author{Krisztian Balog}
\affiliation{University of Stavanger}
\email{krisztian.balog@uis.no}
%

\begin{abstract}
We address the problem of constructing a knowledge base of entity-oriented search intents.  Search intents are defined on the level of entity types, each comprising of a high-level intent category (property, website, service, or other), along with a cluster of query terms used to express that intent.  These machine-readable statements can be leveraged in various applications, e.g., for generating entity cards or query recommendations.  By structuring service-oriented search intents, we take one step towards making entities actionable.  The main contribution of this paper is a pipeline of components we develop to construct a knowledge base of entity intents.  We evaluate performance both component-wise and end-to-end, and demonstrate that our approach is able to generate high-quality data.  
\end{abstract}



\maketitle

\section{Introduction}
\label{sec:intro}

Many information needs behind people's searches revolve around specific entities.
Entities, such as people, organizations, or locations are natural units for organizing information; they can provide not only more focused responses, but often immediate answers~\citep{Pound:2010:AOR}.    
Alongside mere informative exploration, users frequently look for transaction-oriented entity intents, like booking a flight or a hotel.  
In this paper, we propose to build a knowledge base (KB) of entity-oriented search intents.  
Specifically, we identify the main search intents for a representative sample of entity types, and represent them in a structured fashion.  
These machine-readable statements can be used for automatic querying and reasoning about search intents, and represent one step towards making entities \emph{actionable}.

\begin{table}[t]
  \centering  
  \caption{An excerpt from our knowledge base, for IntentID: \texttt{<aviation.airline-65-customer\_service>}.}
  \vspace{-0.75\baselineskip}
  \begin{tabular}{ l l l }
    Predicate & Object & Conf. \\
    \midrule
        \textsc{searchedForType} & Aviation/Airline & 1 \\
        \textsc{ofCategory} & Service & 0.866 \\
        \textsc{expressedBy} & customer service & 0.688 \\
        \textsc{expressedBy} & customer care & 0.656 \\
  \end{tabular}
  \label{table:eg_triples}
  \vspace{-1.\baselineskip}
\end{table}

Most entity-oriented queries consist of an entity name, complemented with context terms to express the underlying intent of the user~\citep{Pound:2010:AOR}.  
Examples of these terms, hereafter named \emph{refiners}, are ``movies'' in the query \emph{``the rock movies''} and ``nyc'' in \emph{``hilton nyc.''}  
We propose to represent search intents at the level of entity types, where a type is a semantic class that groups multiple entities.  
This allows us to capture intents common to many entities, resulting in a representation that is space-efficient and generalizes well to long-tail and emerging entities~\citep{Nakashole:2013:FST,Lin:2012:AOA}.  
We present a pipeline approach that consists of four main components. 
First, we acquire entity-bearing queries of given types and aggregate refiners to obtain type-level query patterns (e.g., ``\emph{[hotel] booking}" or ``\emph{[airline] customer service}").  
Second, we classify each of these type-level refiners into four main \emph{intent categories}, property, website, service, or other, based on how that information need can be fulfilled.  
Third, type-level refiners that express the same underlying intent are clustered together.  
A cluster of refiners corresponds to the various ways of expressing that intent in an actual query (like ``\emph{[hotel] booking}," ``\emph{[hotel] reservations}," ``\emph{[hotel] book a room},'' etc.).
Fourth, structured representations of intents are created by assigning each intent a unique intentID.  
This structured representation then contains the type of the entity, the category of the intent (according to the four main intent categories), and the different ways that intent may be expressed.  
Additionally, each piece of information is assigned a confidence score.  
Table~\ref{table:eg_triples} shows an excerpt from our IntentsKB,  which represents the intent of customer service for an airline.

In summary, this paper makes the following contributions:
\begin{itemize}[leftmargin=*]
	\item We formally define the problem of constructing a knowledge base of entity-oriented search intents, and design a model for structured representation of this knowledge (Sec.~\ref{sec:problem}).
	\item We propose a pipeline framework to build the knowledge base, consisting of refiner acquisition, refiner categorization, intent discovery, and knowledge base construction stages (Sec.~\ref{sec:approach}).
	\item We evaluate the components of our pipeline against editorial judgments, and, using a small seed of labeled data, generate a knowledge base comprising over 30k intents (Sec.~\ref{sec:exper}).
\end{itemize}
The resulting \emph{IntentsKB} knowledge base and the corresponding resources are made available at \url{http://bit.ly/cikm2018-intentsKB}.
\section{Related Work}
\label{sec:rel}

Most related to our paper is the work by \citet{Reinanda:2015:MRR}, who explore entity aspects in user interaction log data.  
Beyond finding aspects by comparing clustering methods over refiners, they address the tasks of ranking such intents for a given entity independently from a query and recommending aspects.  
Unlike them, we (i) work on individual query refiners, (ii) model entity intents at the level of types, (iii) consider always a query context of entities, (iv) use approaches defined entirely in absence of log data, and (v) obtain uniquely identified intents.

\emph{Actions} are a particular kind of entity intent, which we represent using the \emph{service} category.  
The schema.org ontology is equipped with a dedicated actions vocabulary.  
We find this representation well-oriented to model actions, yet not expressive enough regarding the actual ways of expressing the intent in a query.  
Also, the current schema.org model does not allow to represent actions at the entity type level.  
\citet{Lin:2012:AOA} propose the model of active objects as a representation of an entity associated with a ranked list of potential actions.  
These actions are simple surface forms mined from query logs.     
While our envisaged scenario is the same, we differ from them in that we (i) model the spectrum of actions at the level of entity types, (ii) do not work with a fixed set of mined actions but rather discover them automatically from query intents, and (iii) obtain structured triples with uniquely identified actions and their surface forms.
The \emph{action mining} task at the NTCIR Actionable Knowledge Graph track \citep{Blanco:2017:TCE} addresses the problem of ranking potential actions for a given entity of a given type.  Here, an action comprises of a verb and possibly an object or modifier.  While this effort points in the same direction as our work, we note the following limitations:  (i) they consider entity type only as input signals, but the output actions are required at the entity level, (ii) they do not account for actions uniquely identified in a KB, and (iii) they force an action to be expressed as a verbal phrase, which is not the case for most of the service-oriented intents we observed.\footnote{As an example of this last item, \emph{``hilton taxi''} very likely expresses the intent of getting a taxi to the hotel without being the refiner ``taxi'' a verbal phrase.}

\section{Problem Definition}
\label{sec:problem}

Our goal is to build a knowledge base of entity-oriented search intents (\emph{intents} for short), at the level of entity types. 
Each search intent is uniquely identified by an \emph{intentID} and is described by an \emph{intent profile}. The KB consists of a set of (\emph{subject}, \emph{predicate}, \emph{object}, \emph{confidence}) quadruples.  
We formally define the KB as a relational knowledge representation model: $IntentsKB = Q_{T} ~\cup~ Q_{C} ~\cup~ Q_{L}$, 
where the three sets of quadruples are partitioned as follows:

\begin{itemize}[leftmargin=*]
	\item $Q_{T} \subseteq I \times \{$\textsc{searchedForType}$\} \times TYPES \times [0, 1]$;
	\item $Q_{C} \subseteq I \times \{$\textsc{ofCategory}$\} \times CATEGORIES \times [0, 1]$;
	\item $Q_{L} \subseteq I \times \{$\textsc{expressedBy}$\} \times STRINGS \times [0, 1]$.
\end{itemize}
The set $I$ consists of the unique intent identifiers. \textsc{searchedForType}, \textsc{ofCategory}, and \textsc{expressedBy} are predicates used for associating an intent with an entity type, intent category, and possible lexicalizations, respectively. $TYPES$ is the set of entity types in the reference KB, and $CATEGORIES$ is a scheme of intent categories (that is, $\{$Property, Website, Service, Other$\}$, cf. Sect.~\ref{sec:approach:categ}).
\section{Proposed Framework}
\label{sec:approach}

\begin{figure*}[t]
	\centering
	\vspace*{-0.5\baselineskip}
	\includegraphics[width=0.9\textwidth]{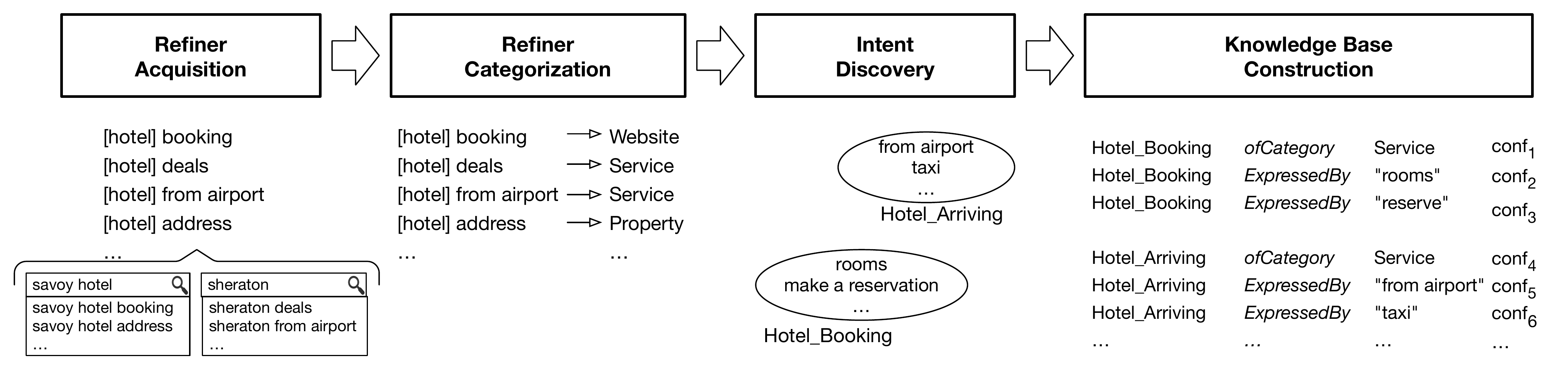}
	\vspace*{-0.75\baselineskip}
	\caption{Overview of our proposed framework. Top: pipeline architecture. Bottom: an example through all its stages.}
	\label{fig:overview}
	\vspace*{-0.75\baselineskip}
\end{figure*}

Figure~\ref{fig:overview} displays an overview of our framework for constructing a knowledge base of entity-oriented search intents.

%
\subsectionshrink
\subsection{Refiner Acquisition}
\label{sec:approach:acqui}

In this first stage, we obtain popular type-level query patterns.  
We sample queries for a set of prominent entities, extract refiners from them, and aggregate these refiners across entity types.  
Specifically, we focus on refiners that complement a mention of a particular entity, i.e., all queries follow the pattern ``\emph{[entity] [refiner]},'' where square brackets indicate placeholders.  
The entity mention is then replaced with the corresponding entity type in the reference knowledge base.  
For example, given the queries ``\emph{sydney map}'' and ``\emph{paris map},'' we extract the type-level query pattern ``\emph{[travel destination] map}.''  
Hereinafter, \emph{type-level refiner}, or simply \emph{refiner}, refers to the query suffix in a type-level query pattern.

\subsectionshrink
\subsection{Refiner Categorization}
\label{sec:approach:categ}

In recent work, we have proposed a suitable scheme to classify entity-related search intents~\citep{Garigliotti:2018:TUE}.
This scheme consists of four \emph{intent categories}, which focus on the way and the type of source to fulfill the information need.
\begin{itemize}[leftmargin=*]
	\item \textbf{Property}: The refiner is about getting a specific entity property or attribute that can be looked up in a knowledge base.  
For example, ``children'' in the query ``\emph{angelina jolie children},'' or ``opening times'' in ``\emph{at\&t stadium opening times}.''  
The criterion does not require the refiner to exist as a property in an actual knowledge base, but rather its existence to be reasonable.
	\item \textbf{Website}: The refiner looks to reach a specific website or application, which is a rough equivalent of navigational queries in~\cite{Broder:2002:TWS}.  
For example, ``twitter'' in the query ``\emph{karpathy twitter}.'' 
	\item \textbf{Service}: The refiner expresses the need to interact with a service, possibly by redirecting to an external site or app.  
For example, ``menu'' in the query ``\emph{keens steakhouse menu}'' would indicate the need to access to an external site for reading the restaurant's menu.  
As another example,  ``new album'' in ``\emph{eric clapton new album}'' looks for a service to read about, or listen to, or buy the new album.  
The interaction would possibly involve further parameters, like ``from'' and ``to'' values for ``ticket price'' in the query ``\emph{jpass bullet train ticket price}.''  
	\item \textbf{Other}: None of the previous ones is applicable.  
For example, ``jr'' in the query ``\emph{tim hardaway jr}'' merely serves to disambiguate the person from other people with the same name.
\end{itemize}
While this categorization comes from our prior work, the automatic assignment of intent categories to refiners has not been addressed to date.  
We approach this task, referred to as \emph{refiner categorization}, as a single-class classification problem using supervised learning.  
Each type-refiner pair defines an instance, and the four categories defined above are the possible classes.

\subsectionshrink
\subsubsection{Lexical features}

We compute attributes at the lexical level of a refiner. 
Many of these signals are detected in the search results obtained from a major Web search engine.  
For every type-refiner pair, we take the most prominent entity assigned to that type (see Sect.~\ref{sec:exper:acqui} for how entity prominence is measured).  
We then search for the corresponding actual entity-bearing query.  
For example, for the type-level query ``\emph{[travel destination] map},'' the corresponding entity-bearing query is ``united states map.'' 
We exploit the top 10 search results to quantify, among others, the size of the set of their URL domains, or, following \citet{Reinanda:2015:MRR}, the average Jaro distance between a refiner and each result URL.
The detailed description of the full set of lexical features can be found in the online repository accompanying this paper.

\subsectionshrink
\subsubsection{Semantic features} 

The semantic similarity between a refiner $r$ and a type $t$ is defined as the cosine similarity $cos(\vect{r}, \vect{t})$ between the centroid word embedding vectors of the refiner terms ($\vect{r}$) and type terms ($\vect{t}$).
This measure captures the compositional nature of words in both the type label and the refiner, which was shown in~\citep{Mikolov:2013:LRC} to be an effective attribute of a phrase.  We use pre-trained word embeddings provided by the \emph{word2vec} toolkit~\citep{Mikolov:2013:DRW}.

\subsectionshrink
\subsection{Intent Discovery}
\label{sec:approach:disco}

Once each type-level refiner is mapped to a category as described above, we proceed to discover the intents underlying those refiners.  We achieve this by clustering the refiners that express the same user intent.
We make use of the intent categories that have been assigned in the previous step, that is, two refiners in the same cluster must have the same intent category.
Following \citep{Reinanda:2015:MRR}, we apply hierarchical agglomerative clustering (HAC) over the distributional semantic space of the refiners for each type.  
As before, we use pre-trained \emph{word2vec} word embeddings.  
For each refiner, we take its vector if the refiner is in the embedding vocabulary, otherwise we assign it the normalized centroid of the vectors of its terms.  

The clusters merging step of HAC is stopped when all the inter-cluster distances are above a certain cut-off threshold.  
This threshold, $\gamma_{c,t}$, is chosen for each combination of intent category $c$ and entity type $t$.  
However, to avoid overfitting, we only learn a single parameter $\epsilon_c$ for each intent category, and then set $\gamma_{c,t}=\epsilon_c M_{c,t}$, where $M_{c,t}$ is the maximum distance between any pair of refiners belonging to intent category $c$ and type $t$.  
This may be seen as a way of normalization, to account for various cluster sizes.  
To find the best $\epsilon_c$ value, we perform a grid search over $[0..1]$ and pick the value that maximizes the evaluation score against the ground truth clusters, available as training data.

\subsectionshrink
\subsection{Knowledge Base Construction}
\label{sec:approach:kb}

In the last step, we construct the full knowledge base representation of intents, i.e., create \emph{intent profiles}.  
Let $i$ denote an intent profile.  
It consists of the set of refiners $R(i)$ that were clustered together.  
Recall that all these refiners have previously been assigned the same intent category, therefore, there exists a single intent category for the profile.
The profile is assigned a unique \emph{intentID}.  
In the interest of readability, it is a concatenation of the entity type, a numerical ID, and the label of the refiner which is closest to the intent centroid.

Recall, that each intent profile has three types of predicates. We define the confidence for each as follows:
\begin{itemize}[leftmargin=*]
	\item \textsc{searchedForType}: since entity type information comes from a curated KB, the confidence in the assigned type is always 1.
	\item \textsc{ofCategory}: we take the average categorization confidence of the refiners in the profile: $\frac{1}{|R(i)|}\sum_{r \in R(i)} \alpha(r)$, where $\alpha(r)$ is the associated confidence score from the intent categorization step.
	\item \textsc{expressedBy}: it is the similarity between the refiner $r$ and the centroid of all refiners in $R(i)$.  
In our case, refiners are represented as embedding vectors and cosine is used to measure their similarity (cf. Sect.~\ref{sec:approach:categ}).
\end{itemize}
We also assign a confidence score to the intent profile itself, by taking a linear mixture of the category confidence $\alpha(c)$ and the average refiner confidence $\alpha(r)$:

\small
\begin{equation}
 \alpha(i) = {{1}\over{2}} \Big ( \alpha(c) + \frac{1}{|R(i)|} \sum_{r \in R(i)} \alpha(r) \Big ) ~.
 \label{eq:conf}
\end{equation}
\normalsize
%

\section{Experimental Evaluation}
\label{sec:exper}

This section presents the evaluation of our approach.

\subsectionshrink
\subsection{Refiner Acquisition}
\label{sec:exper:acqui}

We use the type system of Freebase.  
It is a two-layer categorization system, where types on the leaf level are grouped under high-level domains.  
We focus on popular DBpedia entities to benefit from a larger and more representative selection of information needs.  
Entity popularity is measured by the number of times the entity's English Wikipedia article has been requested.\footnote{Provided by Wikistats: \url{https://dumps.wikimedia.org/other/pagecounts-ez/}.}
We set empirically a popularity threshold of 3,000 page views per article over a span of one year (from June 2015 to May 2016).  
Given a Freebase type, we select it if it covers at least 100 entities with a popularity above the threshold.  
There are a total of 634 such types.

In a second step, we collect query suggestions from the Google Suggestions API for top 1,000 most popular entities per type.  
Then, we obtain type-level query refiners by replacing the entity by its type, as described in Sect.~\ref{sec:approach:acqui}.  
Finally, we retain only those refiners that occur in at least 5 suggestions for the given type.  This leads to a total of 63,148 distinct type-level refiners for 631 types.

\subsectionshrink
\subsection{Refiner Categorization}
\label{sec:exper:categ}

Our dataset consists of 4,490 instances labelled with one of the four intent categories.  
For obtaining the search results exploited by many features, we utilize the Google Search API.
In a preliminary round, we experimented with a variety of classifiers.    
Results are reported only for the best performing classifier, which is Random Forests. 
We used the following parameters: the number of trees is 100 and the maximum depth of the trees is approximately the square root of the feature set used in that setting.  
We train the model using five-fold cross-validation, and report the results in Table~\ref{table:categorization}.  
We find that semantic features perform better than lexical features.  
When combining the two, the resulting performance is inferior to using semantic features alone. Therefore, we only the use semantic feature group in the remainder of our experiments.

\begin{table}[t]
  \small
  \centering
  \caption{Refiner categorization results.}
  \vspace*{-0.75\baselineskip}
  \begin{tabular}{ l c }
	\toprule    
	Feature Set & Accuracy \\
    \midrule
    Feature group I (lexical) & 0.5920 \\ 
    Feature group II (semantic) & \textbf{0.7024} \\ 
    Combined (I + II) & 0.6150 \\
    \bottomrule
  \end{tabular}
  \label{table:categorization}
  \vspace{-0.05in}  
\end{table}
%

\subsectionshrink
\subsection{Intent Discovery}
\label{sec:exper:disco}

We evaluate the clustering approach in two evaluation settings: (i) an ``oracle'' setting, which uses the ground truth intent categories, and (ii) a realistic setting, where the categories are automatically assigned by our refiner categorization component.
We perform five-fold cross-validation for each evaluation instance, with the same partition of folds used for evaluating refiner categorization.  
The training folds are used to optimize $\epsilon_c$ for each category as described in Sect.~\ref{sec:approach:disco}.  
The clusters obtained for a given type are ``flattened'' by ignoring the intent categories, and are evaluated against the ground truth clusters flattened in the same way.  
This serves to eliminate, to some extent, errors that originate from incorrect refiner categorization.
Table~\ref{table:clustering} presents the results.  We find that our automatic setting can achieve a V-measure that is only about 6\% lower than using oracle categories.

\begin{table}[t]
  \small
  \centering
  \caption{Clustering refiners using oracle vs. automatically assigned intent categories, measured in terms of homogeneity, completeness, and V-measure scores.}
  \vspace*{-0.75\baselineskip}
  \begin{tabular}{ l c c c }
	\toprule    
	Category & Hom. & Compl. & V-measure \\
    \midrule
    Oracle categories & 0.9494 & 0.7270 & 0.8201 \\    Automatic categorization & 0.8872 & 0.6872 & 0.7710 \\
    \bottomrule
  \end{tabular}
  \label{table:clustering}
  \vspace{-0.1in}  
\end{table}
%

\subsectionshrink
\subsection{Fact Validation}
\label{sec:exper:facts}

In this last part, we evaluate the end-to-end approach by applying the pipeline, which we trained on 50 types (cf. Sect.~\ref{sec:exper:acqui}), on the remaining 581 types.  As a result, we obtain 31,724 intent profiles that comprise a total of 155,967 quadruples.  
We proceed to estimate its expected overall quality by taking a stratified sample from the generated quadruples.
Considering the confidence scores associated with the profiles (cf. Eq.~\eqref{eq:conf}), we partition the range [0..1] into 5 equally-sized buckets.  
For each bucket, we take 25 random types without repetition and select 5 intent profiles from each type.  
The resulting sample has a total of 2,010 quadruples (around 1.29\% of the size of the KB).
For the following experiment, we ignore the confidence scores, so as not to bias annotators. Thus, we shall refer to triples, not quadruples from now on.

We conduct an annotation experiment to decide whether a triple is correct w.r.t. its intent profile.  
Three expert annotators each manually labeled the sampled triples.  
Note that for every profile, the \textsc{searchedForType} triple is trivially correct.  
Also, if there is only a single \textsc{expressedBy} triple in the profile, then it is trivially correct.  
We measure the inter-annotator agreement using Fleiss Kappa coefficient $\kappa$, separately on \textsc{ofCategory} and \textsc{expressedBy} triples.  
For the set of \textsc{ofCategory} triples, $\kappa = 0.2206$, indicating a fair agreement; for \textsc{expressedBy} triples, $\kappa = 0.8606$, thus it is almost perfect.  
We take the majority vote of the annotators as the ground truth.
Among the \textsc{ofCategory} triples, 88\% of them are correct.  
As for the \textsc{expressedBy} triples, 42\% of them are correct. 
Over all the triples, 54\% of them results to be correct (ignoring the trivially correct \textsc{searchedForType} triples).   

\begin{figure}[!t]
    \centering
    \includegraphics[width=0.42\textwidth]{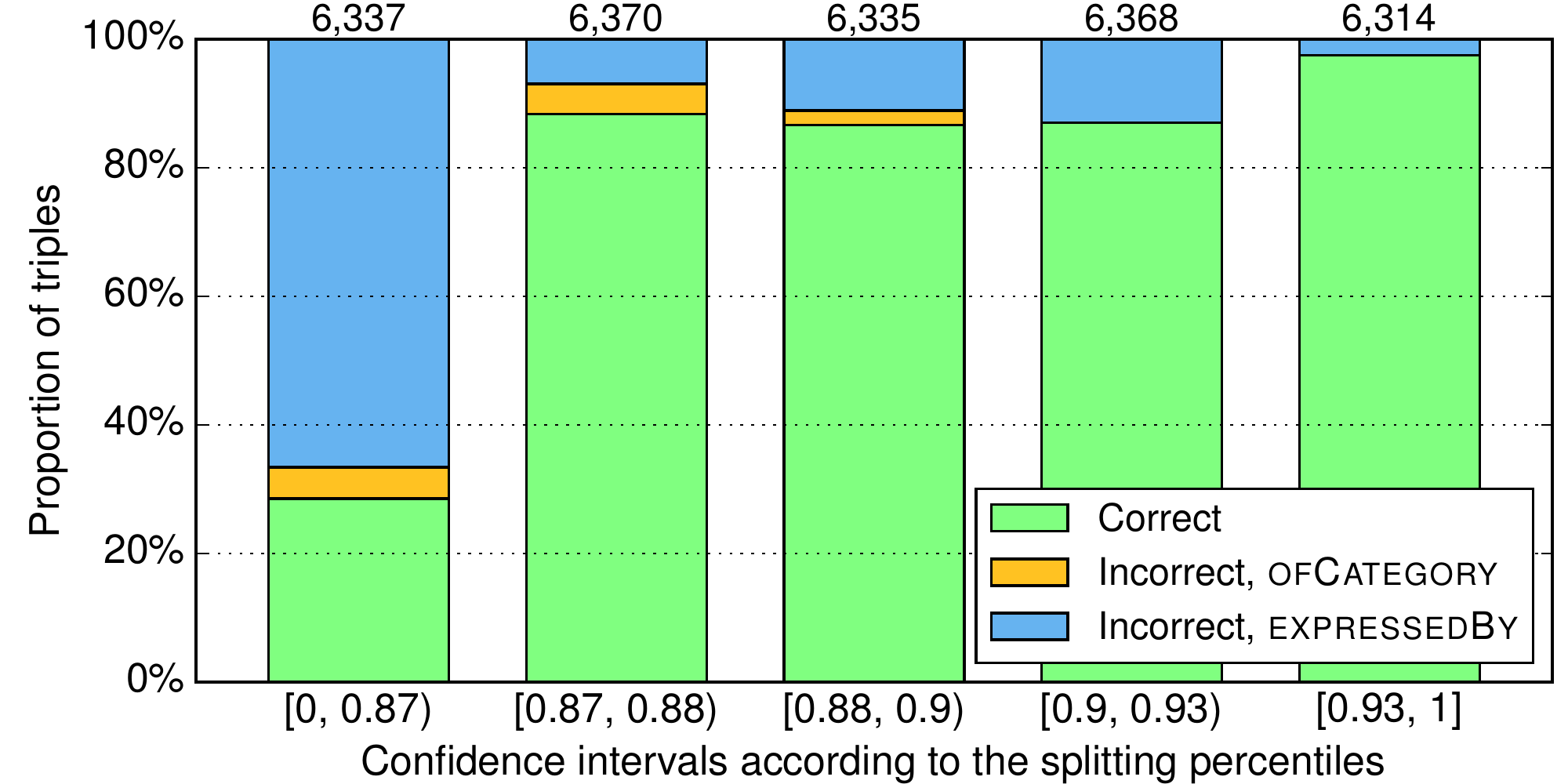}
	\vspace*{-0.5\baselineskip}
    \caption{Proportion of triples in the annotated sample (y-axis), and number of intent profiles in the KB (on top of each bar), per confidence bucket.}
    \label{fig:buckets_vs_triples}
	\vspace*{-0.75\baselineskip}
\end{figure}

How well do the estimated confidence scores correspond to the actual correctness of triples?  
Figure~\ref{fig:buckets_vs_triples} displays the number of correct and incorrect triples with a break down per confidence bucket.
We find that, indeed, the higher the associated confidence score, the more likely it is that the triple is correct.  
It is worth pointing out that accuracy generally is quite high.  
Apart from the leftmost bucket, triples overall have over 90\% accuracy based on the manually labeled sample.

%

\section{Conclusions}
\label{sec:concl}

We have addressed the problem of constructing a knowledge base of entity-oriented search intents and proposed a pipeline approach.    
We have performed an experimental evaluation on the level of individual components as well as on the end-to-end task.
Using a sample of 4.5K labeled instances for 50 types as training data, we have generated a knowledge base of over 30K intents for almost 600 unseen types.  
In future work, we aim to perform extrinsic evaluation, by utilizing our IntentsKB as part of a larger task.

\FloatBarrier  

\bibliographystyle{ACM-Reference-Format}
\bibliography{cikm2018-intents-kb}

\end{document}